\documentclass[11pt]{article}
\usepackage[margin=1in]{geometry}
\usepackage{graphicx}
\usepackage{wrapfig}
\usepackage{hyperref}
\usepackage{eqnarray,amsmath}
\usepackage[symbol]{footmisc}

\def\beq{\begin{equation}}
\def\eeq#1{\label{#1}\end{equation}}
\def\eeqn{\end{equation}}

\def\beqa{\begin{eqnarray}}
\def\eeqa#1{\label{#1}\end{eqnarray}}
\def\eeqan{\end{eqnarray}}

\let\bar=\overbar

\def\Dslash{\not{\hbox{\kern-4pt $D$}}}
\def\dslash{\not{\hbox{\kern-2pt $\del$}}}

\def\msb{{\bar{\ssstyle M \kern -1pt S}}}

\def\Title#1{\begin{center} {\Large {\bf #1} } \end{center}}
\def\Author#1{\begin{center} {\normalsize {\sc #1} } \end{center}}
\def\Institution#1{\begin{center} {\normalsize {\it #1} } \end{center}}
\def\Abstract#1{\noindent {\normalsize {\bf Abstract:} {\normalfont #1}}}
\def\Conference{\vspace{4mm}\begin{raggedright} {\normalsize {\it Talk presented at the 2019 Meeting of the Division of Particles and Fields of the American Physical Society (DPF2019), July 29--August 2, 2019, Northeastern University, Boston, C1907293.} } \end{raggedright}\vspace{4mm}}

\begin{document}

%
%

\Title{Search for exotic decays at NA62}

\Author{Roberta Volpe
  \footnote[2]{roberta.volpe@cern.ch}
  \footnote{On behalf of the NA62 Collaboration: R. Aliberti, F. Ambrosino, R. Ammendola, B. Angelucci, A. Antonelli,
    G. Anzivino, R. Arcidiacono, T. Bache, M. Barbanera,
    J. Bernhard
% ...
}}

\Institution{Centre for Cosmology, Particle Physics and Phenomenology \\ Universit\`{e} Catholique de Louvain, Louvain La Neuve, Belgium}

\Abstract{
  The features of the NA62 experiment at the CERN SPS (high intensity setup, 
  trigger system flexibility, high frequency tracking of beam particles,
  redundant particle identification, and high-efficiency photon vetoes)
  make NA62 particularly suitable to search for long-lived, weakly coupled particles within
  Beyond the Standard Model (BSM) physics, using kaon and pion decays as well as operating the experiment in dump mode.
  The NA62 sensitivity for searches of Dark Photons, Heavy Neutral Leptons and Axion-Like Particles are presented,
  together with prospects for future data taking at the NA62 experiment.
}

\Conference

%
%

\section{Introduction}
\label{sec:intro}
We know that the Standard Model (SM) is not a complete theory,
and the reason why we did not find Beyond SM (BSM) physics up to now could be due to two different situations:
$(i)$ the high energy frontier has not reached the energy scale where the BSM manifests itself;
$(ii)$ the interactions of the new particles with the known physics are too weak to be measured with the LHC luminosity,
for this reason a set of new experiments
is being planned to explore what is called the \textit{intensity frontier}.
A possible situation is that new light particles are coupled to the SM with
small couplings via some
renormalizable or not
interactions, what is known as \textit{portals}.
These portals can be of different nature, vectors, scalars, pseudoscalars,
and can give rise to the production of different kinds of exotic particles,
representing the mediators between the SM and the \textit{hidden sector},
for example Heavy Neutral Leptons (HNL), Dark Photons (DP), Dark Scalars (DS), Axion Like Particles (ALPs) and
these new particles can be in the mass range covered by NA62.
\begin{itemize}
\item {\bf Heavy Neutral Leptons (HNL):}
  The existence of the right handed neutrinos or HNL is one of the most motivated new physics model because
  it can explain several open issues in fundamental physics:
  the light neutrino flavour oscillations,
  the barion asymmetry of the universe, the dark matter.
Mixing of HNL with both $\nu_e$ and $\nu_{\mu}$ can be probed in NA62 by searching for bumps in the missing-mass
distribution of kaons leptonic decays, analysis published ad described in Sec.\ref{sec:HNL_K}.
For larger HNL masses, if they decay to SM particles within the experimental apparatus, searches
in different final states can be performed and the reach is reported in Sec.\ref{sec:HNL_BD}.
\item { \bf Dark photon:} In the minimal version of models with vector portal,
  the mediator is a dark photon which couples to photons via \textit{kinetic mixing} $\epsilon$ \cite{ref:DPtheo}. 
  Ths interaction
  is described by the lagrangian term
  $\epsilon A'_{\mu\nu} F_{\mu\nu}$, where $F_{\mu\nu}$ is the electromagnetic field tensor and $A'$
  is a new vector field corresponding to the mediator called Dark photon. 
  The signature depends on the assumption that the mediator can decay directly to dark matter (DM)
  particles $\chi$ (invisible decays)
  or has a mass below the $2 \cdot m_{\chi}$ threshold and therefore can decay only to SM particles (visible decays).
The first case is considered in a search in pion decays and reported in Sec.\ref{sec:DP_K}.
For $m_{A'} < 2 \cdot m_{\chi}$, $A'$ decays to SM particles and is more convenient searching for it in $D,B$ decays.
Prospects for that are reported in Sec.\ref{sec:DP_BD}.
\item {\bf Dark scalar:}
The mediator between DM and SM particles is 
a light scalar $S$ particle mixing with the Higgs field with the angle $\theta$.
If $S$ decays into dark matter or if its lifetime is long enough to escape the NA62 apparatus, it can be searched for as a peak
in the squared missing mass $m^2_{miss} = (p_K - p_{\pi})^2$, and constitutes a reinterpretation of the main analysis
of NA62 (the study of the SM decay $K^+ \to \pi^+ \nu \bar{\nu}$) which is ongoing \cite{ref:dpf2019proc}.
For $S$ decaying to SM particles within the detector, the search is a bump hunting in 2 lepton invariant mass.
The analysis $K^+ \to \pi^+ S$, with $S \to e^+ e^-$ is ongoing while
the prospects about the search for $B,D \to K S, S \to \mu \mu $ is reported in Sec.\ref{sec:DS_BD}.
\item {\bf Axion Like Particles (ALPs):}
The axion is a very light (sub-eV) exotic particle which was introduced to solve the strong CP problem in QCD.
There are several natural extensions of the axion paradigm, 
which foresee new particles with masses in the MeV-GeV range, called Axion Like Particles (ALPs).
In Sec.\ref{sec:ALP_BD} the sensitivity of NA62 experiment to the ALPS coupled with photons, is reported.
\end{itemize}

\subsection{NA62 beam modes}
High
particle rates are
more easily obtained by fixed target experiments,
hence they are particularly suitable to search for feeble interactions.
Among the already running experiments there is NA62.
Its primary goal is measuring with $BR(K^+\to \pi \nu \bar{\nu})$  with high precision, and it is described shortly in
this same conference proceedings
\cite{ref:myproc} and more extensively in \cite{ref:NA62DP}.
NA62 uses the high intensity 400 GeV proton beam provided by the CERN SPS which impinges a beryllium target
to produce a secondary beam composed, among the other particles, of kaons.
The first period of data taking with all the detectors fully commissioned
and working was in 2016-2018, called NA62Run1 in the following.
After the Long Shutdown 2 (LS2), a new period of data taking is foreseen.
For the $K^+ \to \pi^+ \nu \bar{\nu}$ analysis, a secondary beam of positively charged monochromatic ($\sim$ 75 GeV)
hadrons is selected using a system of magnets and collimators.
The $K^+$, which are only the $6\%$ of the beam, are selected and tracked by dedicated detectors.
With this setup several analyses searching for new exotic particles can be performed and are ongoing.
The results of two of them are reported in Sec.\ref{sec:kaonmode}: the dark photon and the HNL searches.\\
With the standard setup, called \textit{\bf kaon mode} in the following, only new particles with masses lower than
the kaon mass can be searched for.
A different setup, which has already been tested at NA62, consists of closing the aperture of collimators and convert
them in a dump where the protons from SPS are stopped, called in the following \textit{\bf dump mode}.
The interaction of the beam protons with the nuclei of the collimators, which act as dump,
could produce directly new exotic particles, and will produce large samples of SM hadronic particles,
as $D$ and $B$ mesons, which could decay to exotic particles.
The data already collected in dump mode correspond to $2 \times 10^{16}$ proton on target (POT)
and are being analyzed for background studies, while at about $10^{18}$
POT are foreseen to be collected in the next data taking after the LS2. 
In Sec.\ref{sec:BDmode} the sensitivity projections with this amount of data are reported.




\section{Exotic searches with the Kaon beam}
\label{sec:kaonmode}
In the next sections two published results are reported.
They have been performed with the very first datasets collected in kaon beam mode in NA62Run1.
Updates of them and several other analyses with different signatures,
and searching for different hidden sector models, are ongoing.

\subsection{Dark Photon in invisible}
\label{sec:DP_K}
With only $5\%$ of data taken in 2016, a search for DP has been performed in the decay
$K^+ \to \pi^+ \pi^0$, with $\pi^0 \to A' \gamma$.
In the minimal vector portal scenario described in \cite{ref:DPtheo}, if $2 \cdot m_{\chi}<m_{A'}$,
we can assume that $\Gamma(A' \to \bar{\chi} \chi)/\Gamma_{tot} \sim 100\%$.
With this assumption, disregarding if the $A'$ escapes the experimental apparatus without decaying or decays to DM particles,
it is invisible to the NA62 detectors.
The BR is related to the alike SM BR decay by
\begin{equation}
  BR(\pi^0  \to A' \gamma) =  
  2 \cdot \epsilon^2 \Big(1- \frac{M^2_{A'}}{M^2_{\pi^0}} \Big) \cdot BR(\pi^0 \to \gamma \gamma).
\end{equation}
\begin{wrapfigure}{r}{0.6\textwidth}
\centering
\includegraphics[width=0.58\textwidth]{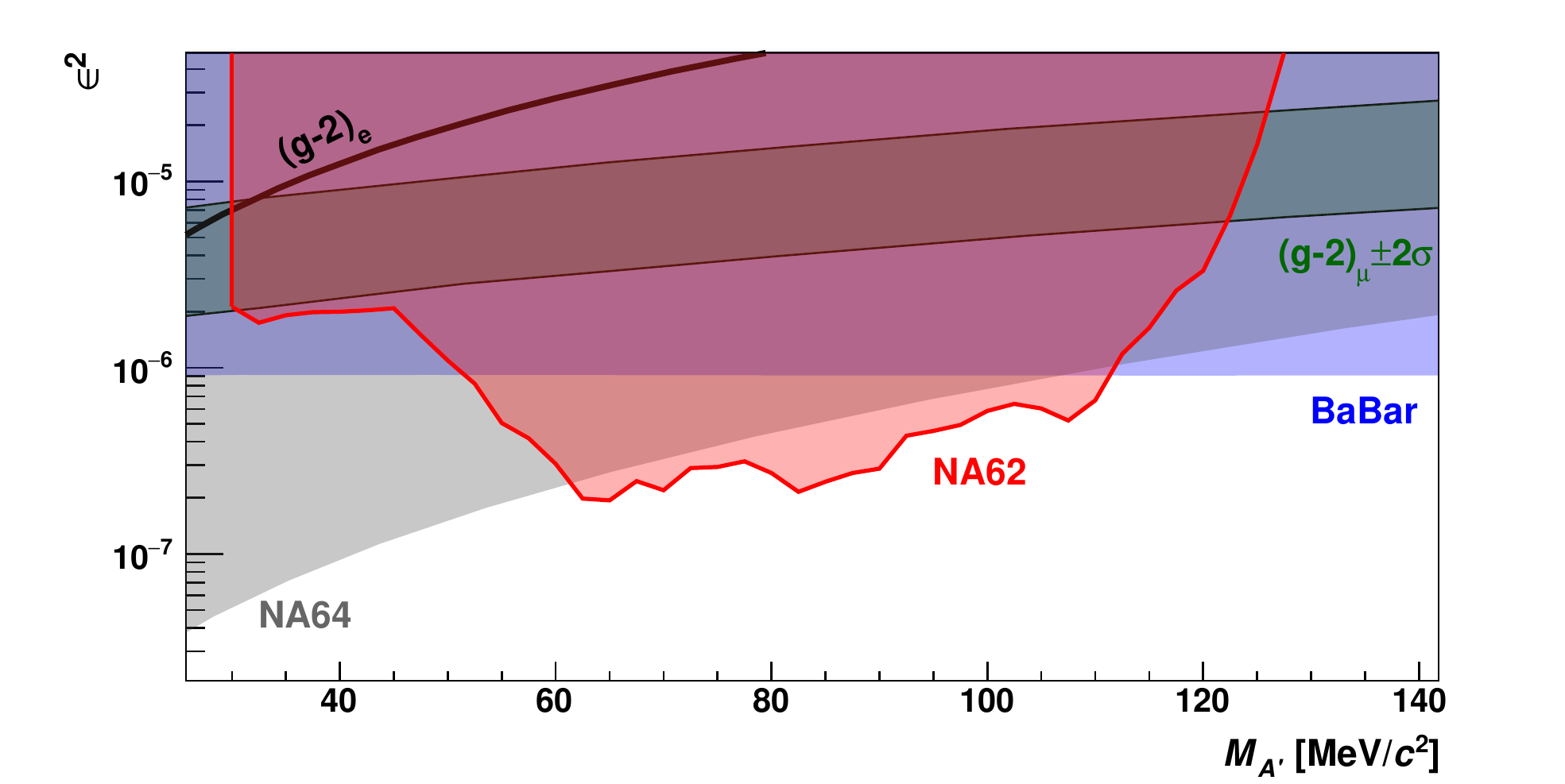}
\caption{Upper limit at 90\% CL from NA62 (red region)
  in the $\epsilon^2$ vs $m_{A'}$ plane with $A'$
decaying into invisible final states.}
\label{fig:DP_K}
\end{wrapfigure}
The analysis is a peak search in the squared missing mass 
$m^2_{miss} = (P_K - P_{\pi} - P_{\gamma})^2$ in the region $m^2_{miss} \in [0.00075, 0.01765 ] $ GeV$^2$/c$^4$,
described in detail in \cite{ref:DP}.
The only non negligible background is due to $\pi^0 \to \gamma \gamma$, with one photon not detected.
This background source distribution has a peak at $m^2_{miss}=0$ and a long tail;
its shape in the signal region is estimated inverting the cut whose inefficiency is responsible
for the photon loss, the normalization is obtained from the distribution at low $m^2_{miss}$, outside the signal region.
No peaks have been found and upper limits have been set to $\epsilon^2$ for $A'$ masses in the mass range $[30,130]$ MeV/c$^2$,
improving on the previous limits over the mass range $[60,110]$ MeV/c$^2$,
as it is shown in Fig.\ref{fig:DP_K}.
The limits from the BaBar(blue) and NA64 (light grey)
experiments are shown as well.
The green band shows the region of the parameter space corresponding
to an explanation of the discrepancy between the measured and expected values of the
anomalous muon magnetic moment $(g-2)_{\mu}$  in terms of a contribution from the $A'$ in the
quantum loops. The region above the black line is excluded by the agreement of the
anomalous magnetic moment of the electron $(g-2)_e$ with its expected value.
After NA62 publication \cite{ref:DP},
NA64 improved the upper limit analyzing the full dataset \cite{ref:NA64}.

\subsection{Heavy neutral leptons}
\label{sec:HNL_K}
A search for HNL from Kaon decays has been performed.
The decay considered is 
$K^+ \to  l^+ N$, with $N$ the HNL and $l=e,\mu$.
If the HNL does not decay to SM particles within the apparatus, the analysis proceeds searching for a peak in the squared missing
mass $m^2_{miss} = (p_{K} - p_{l})^{2} $.
\begin{wrapfigure}{r}{0.6\textwidth}
  \centering
  \includegraphics[width=7cm]{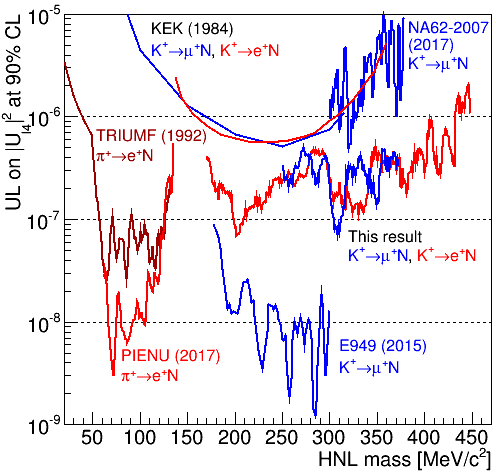}
  \caption{Upper limits at 90\% CL on $|U|^2$
    compared to
    the limits established by earlier HNL production searches in $\pi^+$ decays and $K^+$ decays.}
\label{fig:HNL_K}
\end{wrapfigure}
A first result has been obtained with a small data sample corresponding to $3 \cdot 10^{8}$ $K^+$ decays,
collected in 2015 when the experimental apparatus was not fully commissioned.
In particular the Kaon Tracker (GTK) was not used, this resulted in a $m_{miss}^2$ resolution and signal to background ratio worse
than the potential reach.
Despite that, results competitive with earlier experiments have been obtained.
No signal has been observed.
Upper limits on the mixing coupling with $e$ and $\mu$ SM neutrinos have been set and shown in Fig.\ref{fig:HNL_K}
The result improves the existing limits on $|U_e|$ over the whole mass range considered, [170,448] MeV/c$^2$,
and on $|U_{\mu}|$ above 300 MeV/c$^2$.
New results on this search, obtained using the 2017 dataset, will be presented at KAON2019 conference and will show
large improvements, 
possible thanks to the increased statistics and the full use of the Kaon tracker.

\section{Exotic searches in beam dump mode}
\label{sec:BDmode}
Studies on the hidden sector searches, which can be performed in the next 5,10,15 years with existent and
proposed experiments, are described in the CERN Physics Beyond Collider (PBC), BSM working group \cite{ref:PBC}.
Here a summary on the 5 year timescale, where NA62 is competitive, is reported.
The mediators of the hidden sector models described in Sec.\ref{sec:intro} can be 
produced in proton-nucleus interactions and 
in B and D decays together with a SM particles.
The latter will be stopped or bent by the dump, collimators and the magnets, while the neutral,
feebly interacting exotic particles will proceed straight and are predicted to have displaced vertex decays. 
If the new particle will decay within the NA62 apparatus in visible SM particles,
searches can be performed in the invariant mass distribution of such particles.
An overview of the NA62 reach, for benchmark models where it is competitive in the next 5 years,
is given in the next sections.
It is assumed that $10^{18}$ POT will be collected in the run starting in 2021 after the LS2.
This amount of data can be obtained in three months of data taking at nominal proton beam intensity.
The curves shown refer to 90\% CL upper limit, which 
can be interpreted as
$3\sigma$ discovery in case the backgrounds are maintained below a fraction of event.

\subsection{Heavy neutral leptons}
\label{sec:HNL_BD}
Sensitivity studies on searches for  
heavy neutrinos mixed with $\nu_e, \nu_{\mu}, \nu_{\tau}$ , have been performed considering their decays to SM particles.
If the HNLs exist, they would be produced in every meson decay
containing active neutrinos with a branching fraction
proportional to the mixing parameters $|U_e|,|U_{\mu}|,|U_{\tau}|$.
Within NA62 a study based on HNL produced in D meson decays was performed.
\begin{wrapfigure}{r}{0.5\textwidth}
  \centering
  \includegraphics[height=5cm]{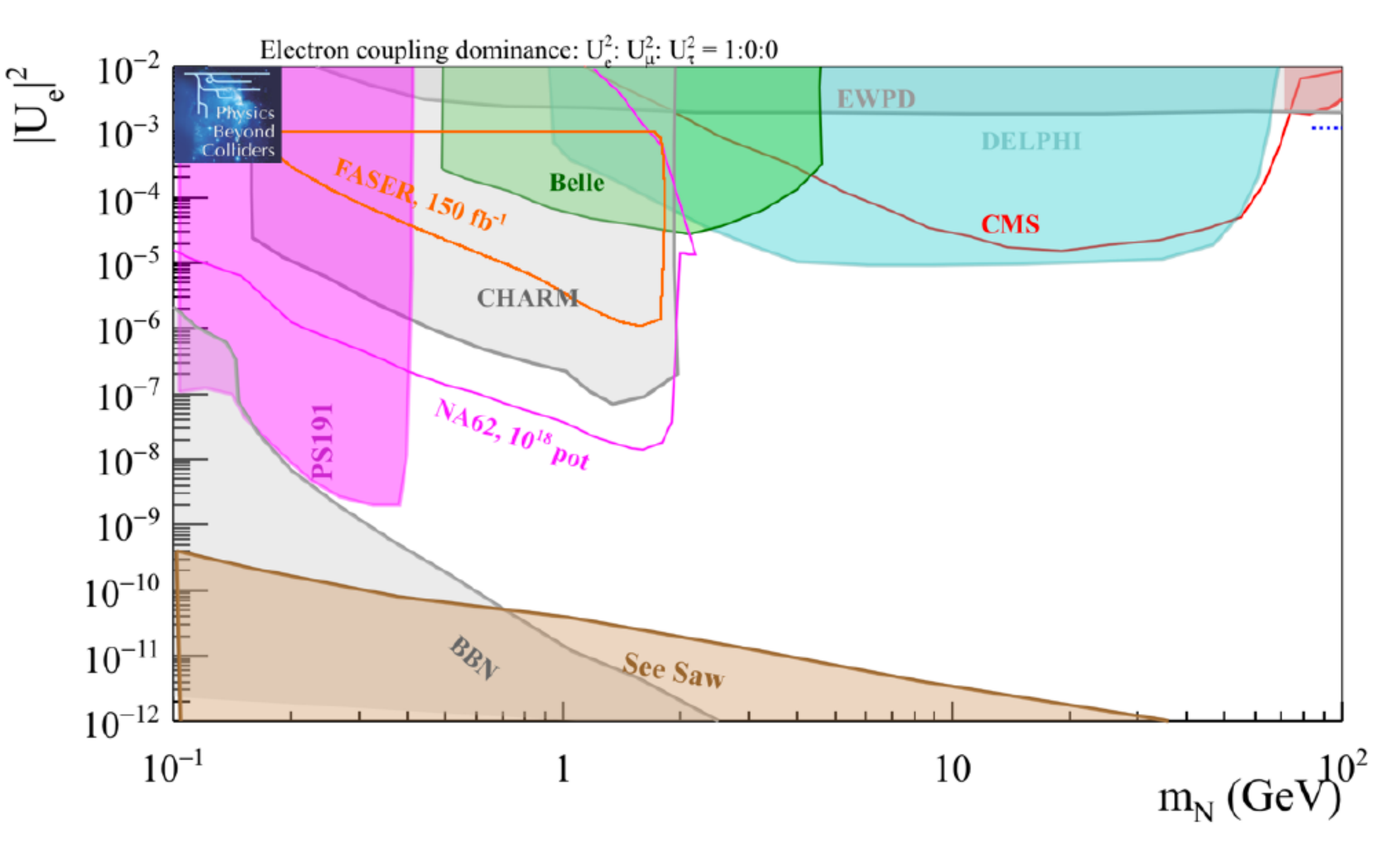}
  \caption{Sensitivity to Heavy Neutral Leptons with coupling
to the first lepton generation only.}
\label{fig:BD_HNLe}
\end{wrapfigure}
The HNL decays considered are $N\to \pi e$ and  $N\to \pi \mu$.
Acceptance, trigger and selection efficiencies were included and zero-background was assumed.
Studies performed with the already acquired $3 \cdot 10^{16}$ POT dataset in dump mode show that
the background can be reduced to zero with the current setup for fully reconstructed final states,
while for open final states the addition of an Upstream Veto in front of the decay volume is required.
Sensitivity to PBC benchmark scenarios in which a HNL couples to one SM generation at the time has been evaluated.
The result of this study is reported, together with other PBC experiments sensitivities,
in Fig.\ref{fig:BD_HNLe} and Fig.\ref{fig:BD_HNL}.
It is clear that within 5 years, before new dedicated facilities will be constructed,
NA62 will be the leading experiment in HNL searches below the $D$ meson mass.
\begin{figure}[htb]
\centering
\includegraphics[height=5cm]{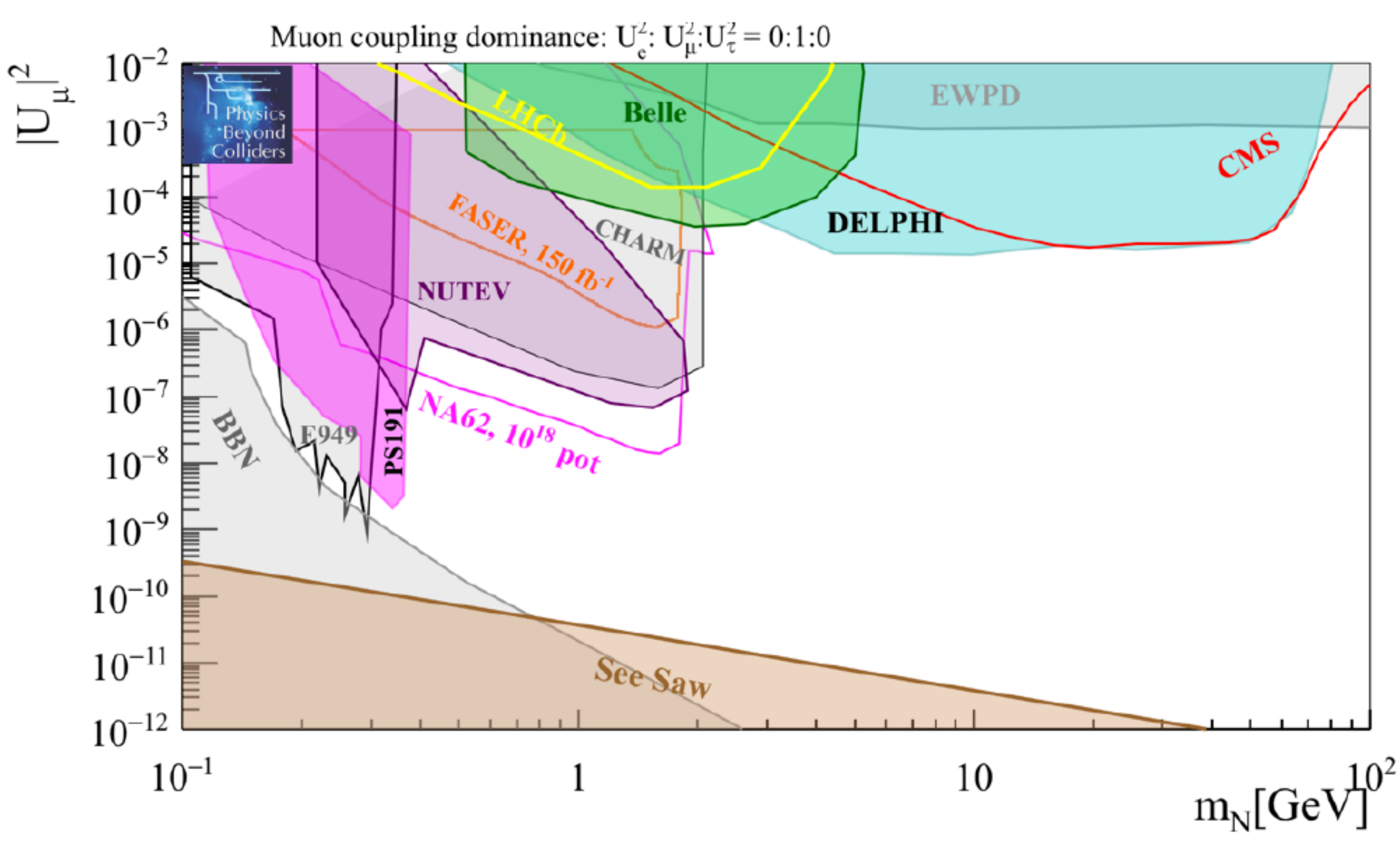}
\includegraphics[height=5cm]{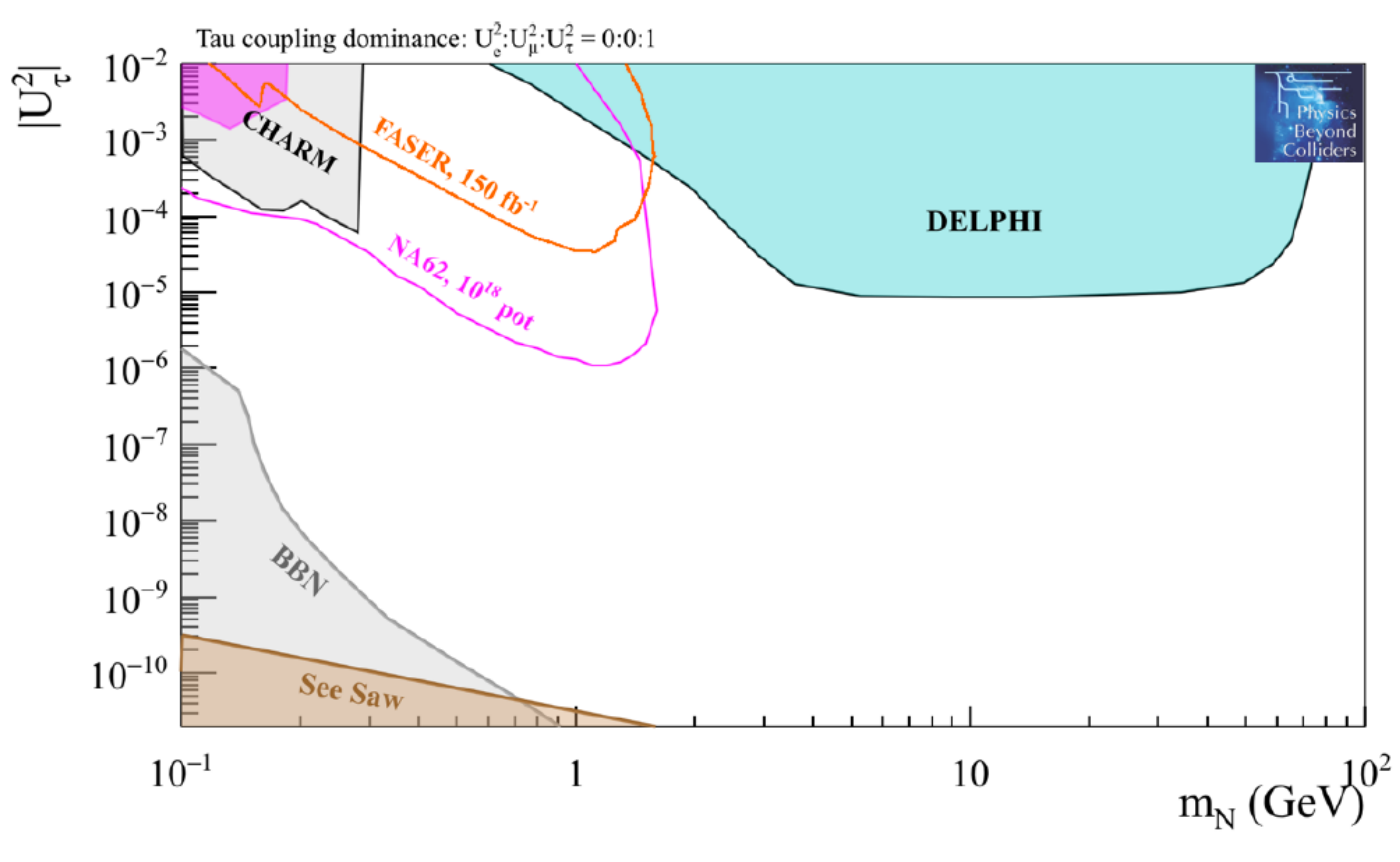}
\caption{Sensitivity to Heavy Neutral Leptons with coupling
to the second (left) and third (right) lepton generation only.} 
\label{fig:BD_HNL}
\end{figure}

\subsection{Dark Photon}
\label{sec:DP_BD}
If the Dark photon exists, it is produced in $B$ and $D$ decays together with other SM particles.
The SM particles will be stopped or bent by the dump, collimators and magnets, while the DP proceeds straight till its decay.\\
\begin{wrapfigure}{r}{0.65\textwidth}
\centering
\includegraphics[width=0.6\textwidth]{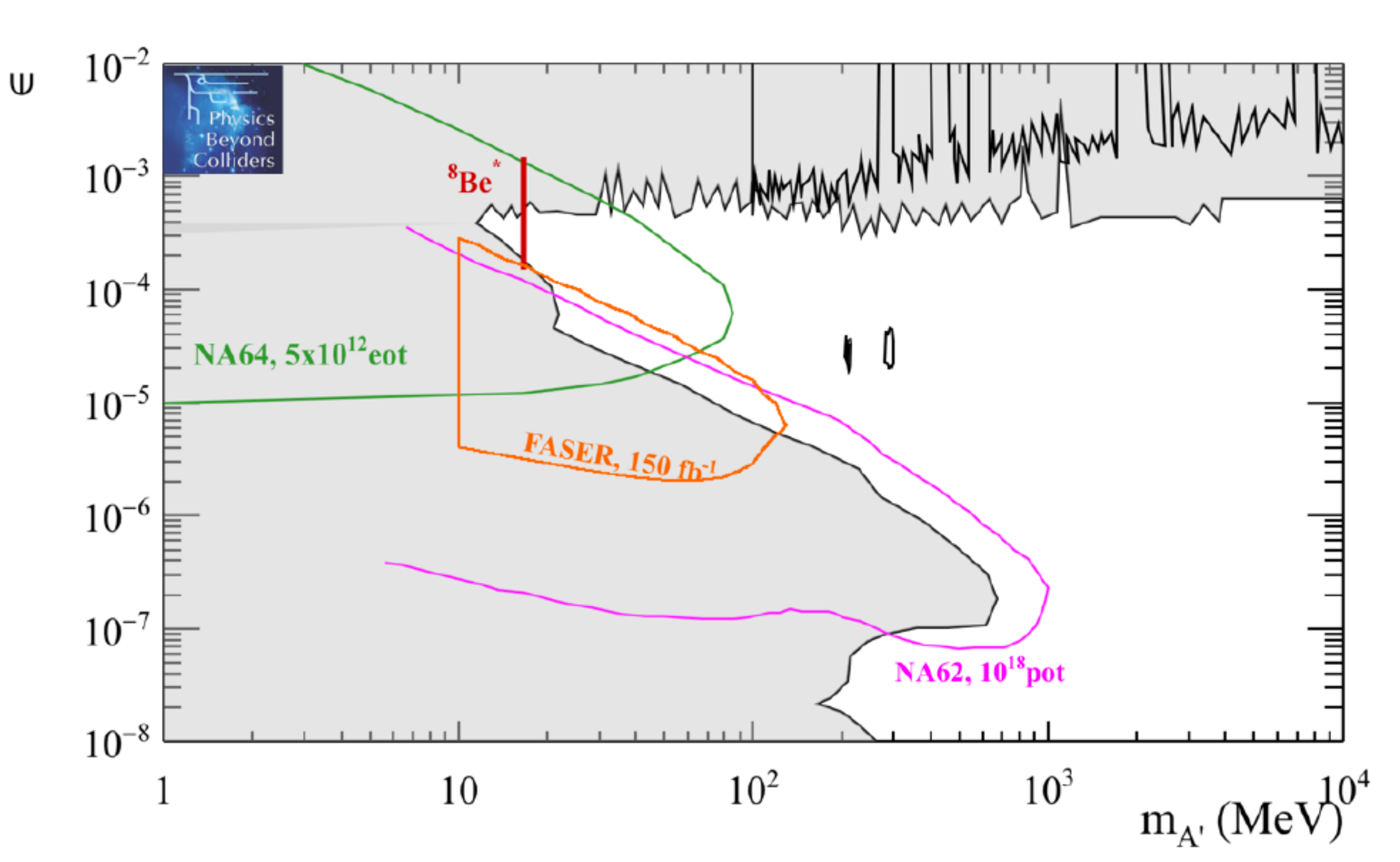}
\caption{Upper limits at 90\% CL PBC projects on $\sim 5$ year timescale.}
\label{fig:BD_DP}
\end{wrapfigure}
If $m_{A'}< 2 \cdot m_{\chi}$, it can decay only to SM particles.
In this study the decays $A'\to e e$ and $A'\to \mu \mu$ have been looked at,
taking into account acceptance, trigger and selection efficiencies.
The study was performed with a toy-MC and 
cross-checked against the full MC.
Fig.\ref{fig:BD_DP} shows the $90\%$ CL upper limit obtained
in the plane of mixing strength $\epsilon$ versus mass $m_{A'}$.
This projection can be improved taking into account also the $A'$ direct production in the dump via QCD processes and in decays of mesons produced in the dump.


\subsection{Dark scalars}
\label{sec:DS_BD}
In the scenario with a dark scalar (DS) mixing with the Higgs with an angle $\theta$,
the lagrangian term describing this interaction is $- (\mu S + \lambda S^2) H^\dag H $, where $S$ and $H$ are the fields for the DS and the Higgs.
  \begin{figure}[hbt]
  \centering
  \includegraphics[height=7cm]{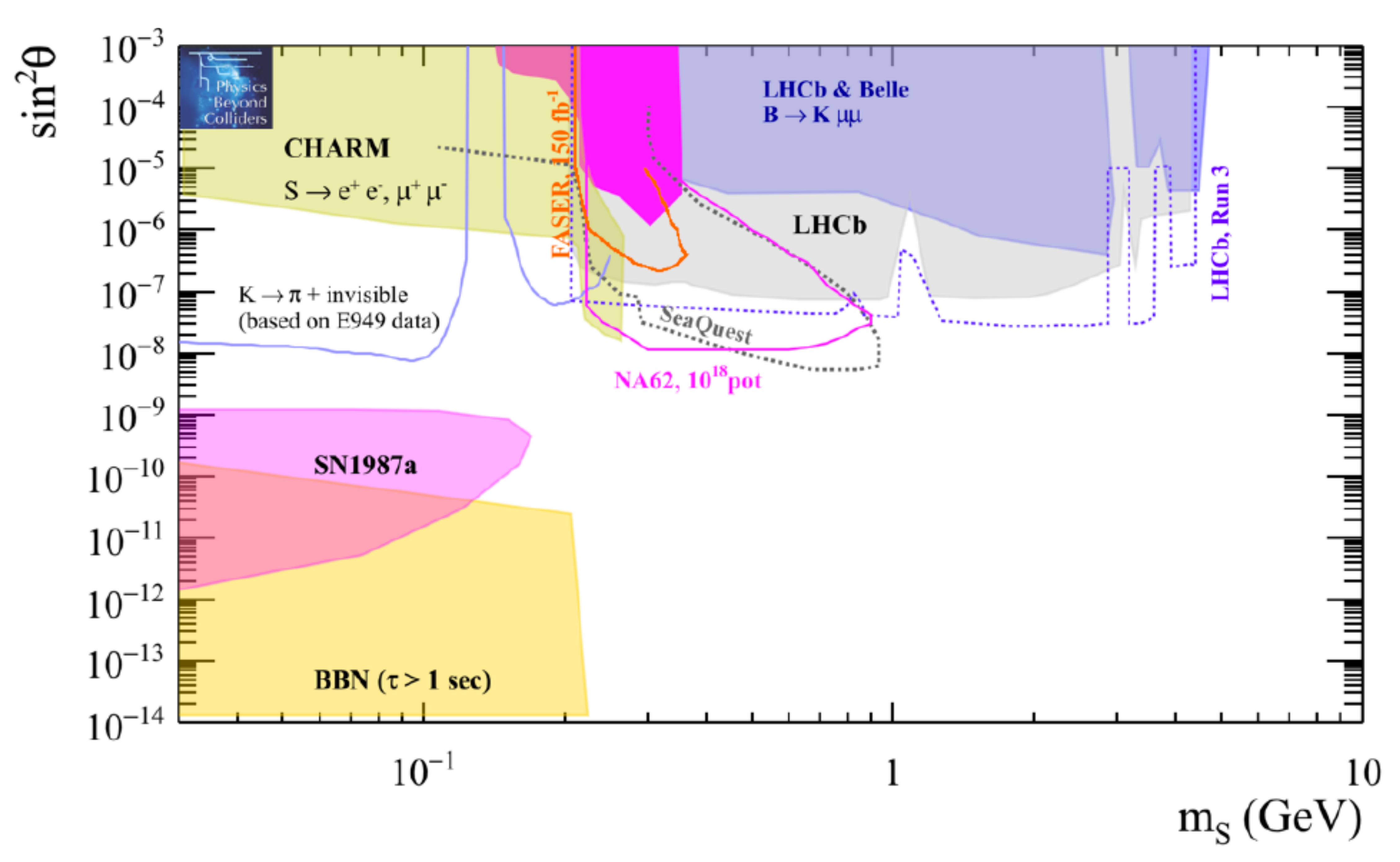}
  \caption{Prospects on $\sim 5$ year timescale for PBC projects for the dark scalar mixing with the Higgs
    in the plane mixing angle $sin^2 \theta$ versus dark scalar mass $m_S$.}
\label{fig:BD_DS}
  \end{figure}
The NA62 studies considered the simple case
in which $\lambda =0$ and $\mu = sin^2 \theta $ for all the production and decay processes. 
The dominant DS production mode is from decays of B mesons which are
produced in the
dump.
The $S$ decays considered are $S \to \mu\mu, ee, \pi \pi, KK$.
The study has been carried out have with a toy-MC and cross-checked with a full MC simulation.
Fig.\ref{fig:BD_DS} shows that NA62 in dump mode 
between the di-muon mass and $\sim 1$ GeV range and will compete with SeaQuest in the same timescale.\\
In addition Fig.\ref{fig:BD_DS} illustrates that for $m_S$ below the kaon mass,
the best sensitivity is obtained with $K^+ \to \pi^+ S$ decay, with $S$ invisible.
The current upper limit has been set by E949 experiment and will be soon improved by NA62
in kaon mode with a reinterpretation of the $K^+\to \pi^+ \nu \bar{\nu}$ analysis.
In the distribution of $m^2_{miss} = (P_{K^+}-P_{\pi^{+}})^{2}$,
the SM process $K^+\to \pi^+ \nu \bar{\nu}$ contributes to the background
with a smooth distribution (see Fig.3 of \cite{ref:myproc})
while the $K^+ \to \pi^+ S$ would result in a peak centered in the $m_{S}$ mass.
With a similar analysis very light (sub-keV) pseudoscalars (for example \cite{ref:axiflavon})
can be searched for with $m^2_{miss}$ centered at 0.



\subsection{Axion Like Particles}
\label{sec:ALP_BD}
Within the PBC study, the NA62 experiment provided the sensitivity to the ALPs coupled to photons.
ALPs can couple to any SM particle, but
if the dominant coupling is with photons,
the ALPs are created mainly via the Primakov production
in the target \cite{ref:ALPS}.\\
\begin{wrapfigure}{r}{0.5\textwidth}
\centering
\includegraphics[width=9cm]{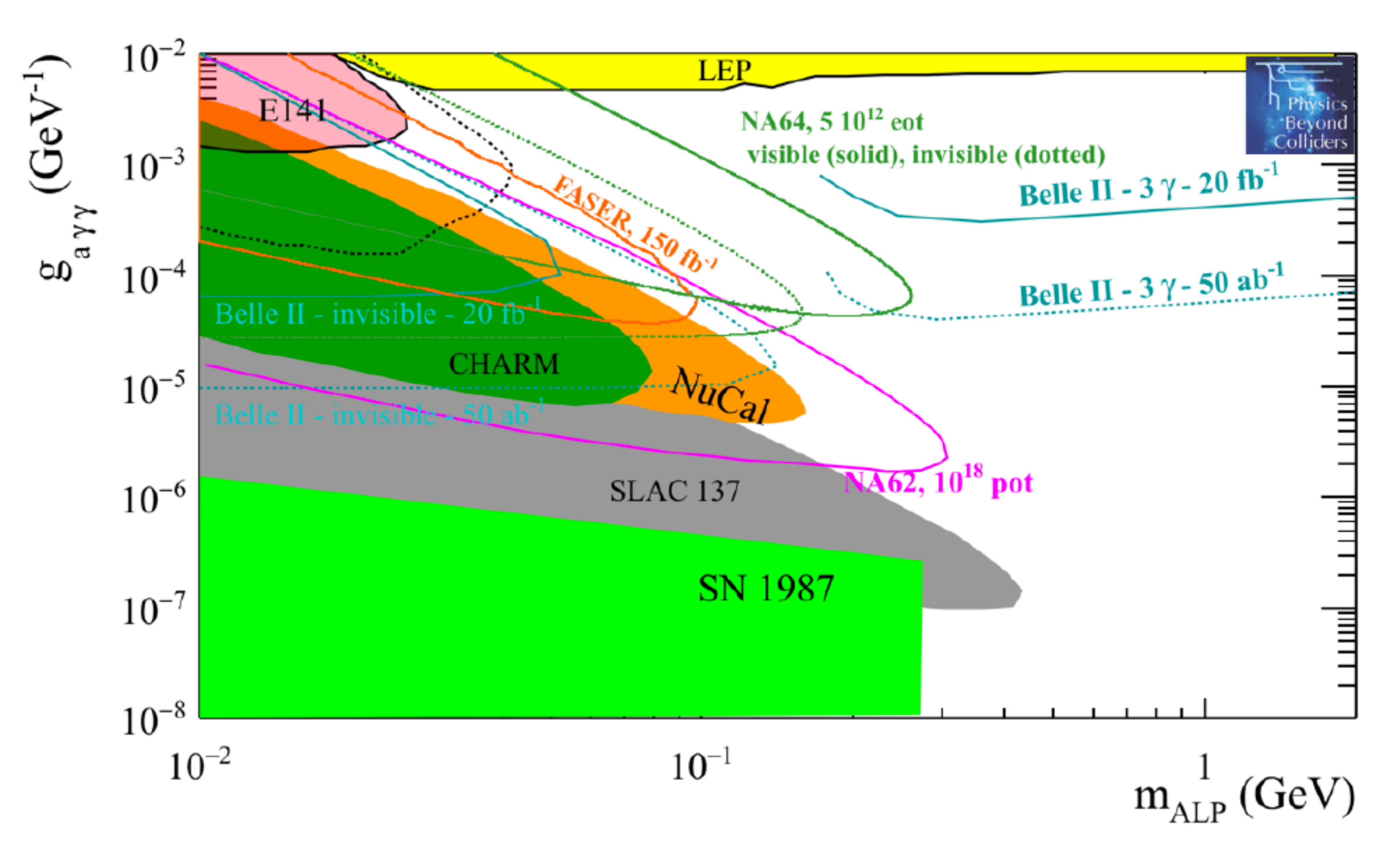}
\caption{ALPs with photon coupling in the plane coupling $g_{a\gamma\gamma}$ versus mass $m_{ALP}$.}
\label{fig:BD_ALP}
\end{wrapfigure}
The term of the Lagrangian describing the photon-ALP interaction is: 
\begin{equation}
L = - \frac{1}{4} g_{a\gamma\gamma} a F^{\mu\nu} \tilde{F}_{\mu\nu} 
\end{equation}
where $F^{\mu\nu}$ is the electromagnetic field and $g_{a\gamma\gamma}$ is the ALP-photon coupling.\\
Fig.\ref{fig:BD_ALP} shows that NA62 can provide the best sensitivity in 5 years
for ALPS decaying into two photons.
The shape of the sensitivity curve, in particular the worse reach at very small masses,
is explained by considering the ALP lifetime.
Indeed for small couplings and masses it can decay outside the experimental
apparatus and in this case no photons can be detected.
This result can be improved taking into account another ALP production mechanism: 
pions and other mesons dominantly decaying in two photons \cite{ref:ALPSnew}.

\section{Conclusions}
Despite it has been built with the aim 
to search for new physics at high scale in indirect way,
the NA62 experiment represents an extraordinary opportunity to search for exotic particles with masses in the range of MeV-GeV,
before dedicated new facilities are built.
With the kaon mode runs, regions of space parameters for dark photon and heavy neutral leptons has been excluded
and updates to extend those regions are expected very soon with the analysis of the full NA62Run1 dataset.
The analysis of the NA62Run1 dataset in kaon mode will cover also a region in the space parameter of the scalar sector.
With a change in the beam line, NA62 turns to a beam dump experiment,
able to produce large samples of B and D mesons and possibly exotic particles.
In the beam dump mode, simulation studies have demonstrated that, within the next 5 years,
NA62 can deliver the best sensitivity for some space parameter ranges in HNL, Dark photon, dark scalar and ALPs searches.   

\section*{Acknowledgements}
R.Volpe was supported by FRS-FNRS under the Excellence of Science (EoS) project n.30820817
be.h \textit{The H boson gateway to physics beyond the Standard Model}.

\end{document}